\documentclass[preprint,floats,aps,epsfig,nofootinbib,amssymb,showpacs]{revtex4}
\usepackage{color}
\usepackage{graphicx}
\usepackage{tabularx}
\usepackage{amsmath}
\usepackage{amssymb}
\usepackage{datetime}
\usepackage[left]{lineno}
\usepackage{txfonts}
\usepackage{dcolumn}
\usepackage{bm}

\begin{document}

\title{Detecting electron neutrinos from solar dark matter annihilation by JUNO}

\author{Wan-Lei Guo}
\email[Email:]{guowl@ihep.ac.cn}

\affiliation{Institute of High Energy Physics, Chinese Academy of
Sciences, P.O. Box 918, Beijing 100049, China }

\begin{abstract}

We explore the electron neutrino signals from light dark matter (DM)
annihilation in the Sun for the large liquid scintillator detector
JUNO. In terms of the spectrum features of three typical DM
annihilation channels $\chi \chi \rightarrow \nu \bar{\nu},\tau^+
\tau^-, b \bar{b}$, we take two sets of selection conditions to
calculate the expected signals and atmospheric neutrino backgrounds
based on the Monte Carlo simulation data. Then the JUNO
sensitivities to the spin independent DM-nucleon and spin dependent
DM-proton cross sections are presented. It is found that the JUNO
projected sensitivities are much better than the current spin
dependent direct detection experimental limits for the $\nu
\bar{\nu}$ and $\tau^+ \tau^-$ channels. In the spin independent
case, the JUNO will give the better sensitivity to the DM-nucleon
cross section than the LUX and CDMSlite limits for the $\nu
\bar{\nu}$ channel with the DM mass lighter than 6.5 GeV. If the
$\nu \bar{\nu}$ or $\tau^+ \tau^-$ channel is dominant, the future
JUNO results are very helpful for us to understand the tension
between the DAMA annual modulation signal and other direct detection
exclusions.

\end{abstract}

\pacs{95.35.+d, 95.55.Vj, 13.15.+g}

\maketitle

\section{Introduction}

The existence of dark matter (DM) is by now well confirmed
\cite{DM1,DM2}. The current cosmological observations have helped to
establish the concordance cosmological model where the present
Universe consists of about 69.3\% dark energy, 25.8\% dark matter
and 4.9\% atoms \cite{Ade:2015xua}. Understanding the nature of dark
matter is a prime open problem in particle physics and cosmology.
Here we focus on the neutrino signals from the DM annihilation in
the Sun. As well as in the DM direct detection experiments, the halo
DM particles can also elastically scatter with nuclei in the Sun.
Then they may lose most of their energy and are captured in the Sun
\cite{DM1}. These trapped DM particles will be accumulated in the
core of the Sun due to repeated scatters and the gravity potential.
Therefore the Sun is a very interesting place for us to search the
DM annihilation signals. Due to the interactions of the DM
annihilation products in the Sun, only the neutrino can escape from
the Sun and reach the Earth. Therefore the terrestrial neutrino
detectors can detect these neutrinos. The Cherenkov detectors
Super-Kamiokande \cite{SK}, IceCube \cite{IC} and ANTARES
\cite{ANTARES} have presented their results.

The Jiangmen Underground Neutrino Observatory (JUNO) \cite{JUNO}, a
20 kton multi-purpose underground liquid scintillator (LS) detector,
are constructing in China to primarily determine the neutrino mass
hierarchy by detecting reactor antineutrinos. The JUNO central
detector as a LS calorimeter has an excellent energy resolution and
a very low energy threshold. It is found that the LS detector has
capability to reconstruct the track direction of the energetic
charged particle by use of the timing pattern of the first-hit on
the photomultiplier tubes (PMTs) since the energetic particle
travels faster than light in the LS \cite{Learned:2009rv,LENA}.
Therefore the JUNO LS detector can detect the neutrinos from the DM
annihilation in the Sun. In Ref. \cite{JUNO}, the JUNO has
calculated the $\nu_\mu/\bar{\nu}_\mu$ signals from the DM
annihilation in the Sun. Some authors have analyzed the
$\nu_e/\bar{\nu}_e$ and $\nu_\mu/\bar{\nu}_\mu$ signals for the
other LS detectors \cite{Kumar}. Here we shall discuss the
$\nu_e/\bar{\nu}_e$ signals from the solar DM annihilation in JUNO.

In this paper, we shall explore the $\nu_e/\bar{\nu}_e$ signals in
JUNO from the light DM annihilation $4 \,{\rm GeV} \leq m_D \leq 20
\,{\rm GeV}$ and consider three typical DM annihilation channels
$\chi \chi \rightarrow \nu \bar{\nu},\tau^+ \tau^-, b \bar{b}$.  Two
sets of selection conditions will be chosen for the monoenergetic
($\nu \bar{\nu}$ channel) and continuous ($\tau^+ \tau^-$ and $ b
\bar{b}$ channels) spectrum cases. Then we calculate the
corresponding selection efficiencies and the atmospheric neutrino
background based on the Monte Carlo (MC) simulation data. The JUNO
sensitivities to the spin-independent (SI) DM-nucleon  and
spin-dependent (SD) DM-proton elastic scattering cross sections will
be given for the three DM annihilation channels. This paper is
organized as follows: In Sec. II, we outline the main features of
the DM captured by the Sun and give the produced neutrino fluxes. In
Sec. III, we present the selection conditions and numerically
calculate the corresponding $\nu_e/\bar{\nu}_e$ event numbers in
JUNO. In Sec. IV, we analyze the expected atmospheric neutrino
background and calculate the JUNO sensitivities to the DM direct
detect cross sections. Finally, a conclusion will be given in Sec.
V.


\section{Neutrinos from the DM annihilation in the Sun}

A halo DM particle via elastic scattering with the solar nuclei may
lose most of its energy and is trapped by the Sun \cite{DM1}. On the
other hand, each DM annihilation in the Sun will deplete two DM
particles.  The evolution of the DM number $N$ in the Sun can be
written as \cite{Griest:1986yu}:
\begin{eqnarray}
\dot{N} = C_\odot  - C_A N^2 \;, \label{N_DM}
\end{eqnarray}
where the dot denotes differentiation with respect to time. The DM
solar capture rate $C_\odot$ in Eq. (\ref{N_DM}) is proportional to
the DM-nucleon (DM-proton) elastic scattering cross section
$\sigma^{\rm SI}_n$ ($\sigma^{\rm SD}_p$) in the SI (SD) interaction
case. In the next paragraph, we shall give the corresponding
formulas to calculate $C_\odot$. The last term $C_A N^2$ in Eq.
(\ref{N_DM}) controls the DM annihilation rate in the Sun. The
coefficient $C_A$ depends on the the thermally averaged annihilation
cross section times the relative velocity $\langle \sigma v \rangle$
and the DM distribution in the Sun. To a good approximation, one can
obtain $C_A = \langle \sigma v \rangle/V_{eff}$, where $V_{eff} =
5.8 \times 10^{30} \; {\rm cm^3} ( 1 {\rm GeV}/{m_D} )^{3/2}$ is the
effective volume of the core of the Sun
\cite{Griest:1986yu,Gould:1987ir}. In Eq. (\ref{N_DM}), we have
neglected the evaporation effect since this effect is very small
when the DM mass $m_D \gtrsim 4$ GeV \cite{Gould:1987ju}. One can
easily solve the evolution equation and derive the DM solar
annihilation rate \cite{Griest:1986yu}
\begin{eqnarray}
\Gamma_A = \frac{1}{2} C_A N^2 = \frac{1}{2} C_\odot \tanh^2
(t_\odot \sqrt{ C_\odot C_A} ) \;, \label{GammaA}
\end{eqnarray}
where $t_\odot \simeq 4.5$ Gyr is the solar age. If $t_\odot \sqrt{
C_\odot C_A} \gg 1$, the DM annihilation rate reaches equilibrium
with the DM capture rate. In the equilibrium case, one may derive
the maximal DM annihilation rate $\Gamma_A = C_\odot/2$ which means
that the DM annihilation signals will only depend on $\sigma^{\rm
SI}_n$ or $\sigma^{\rm SD}_p$.

\begin{figure}[htb]
\begin{center}
\includegraphics[scale=0.46]{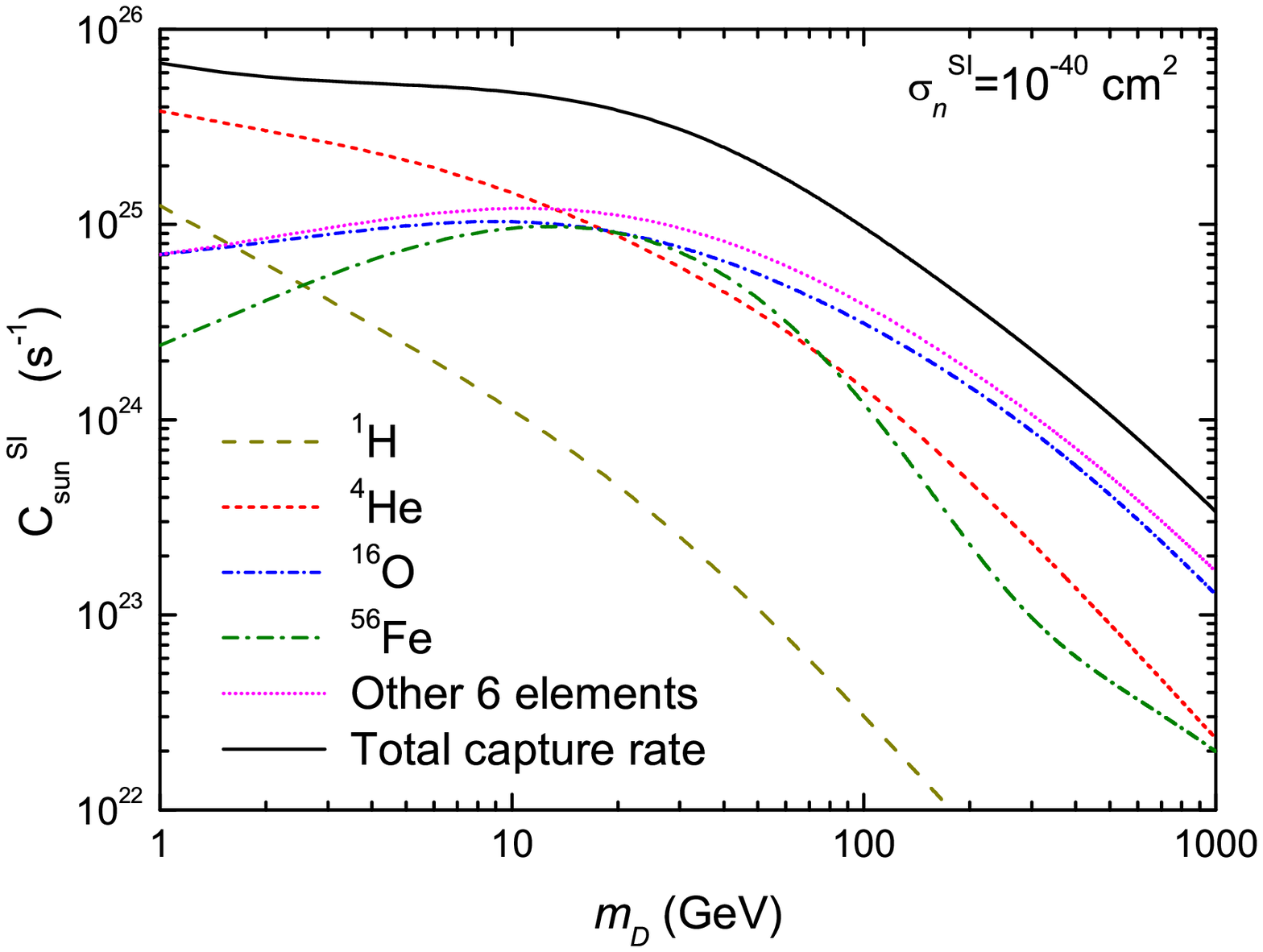}
\includegraphics[scale=0.46]{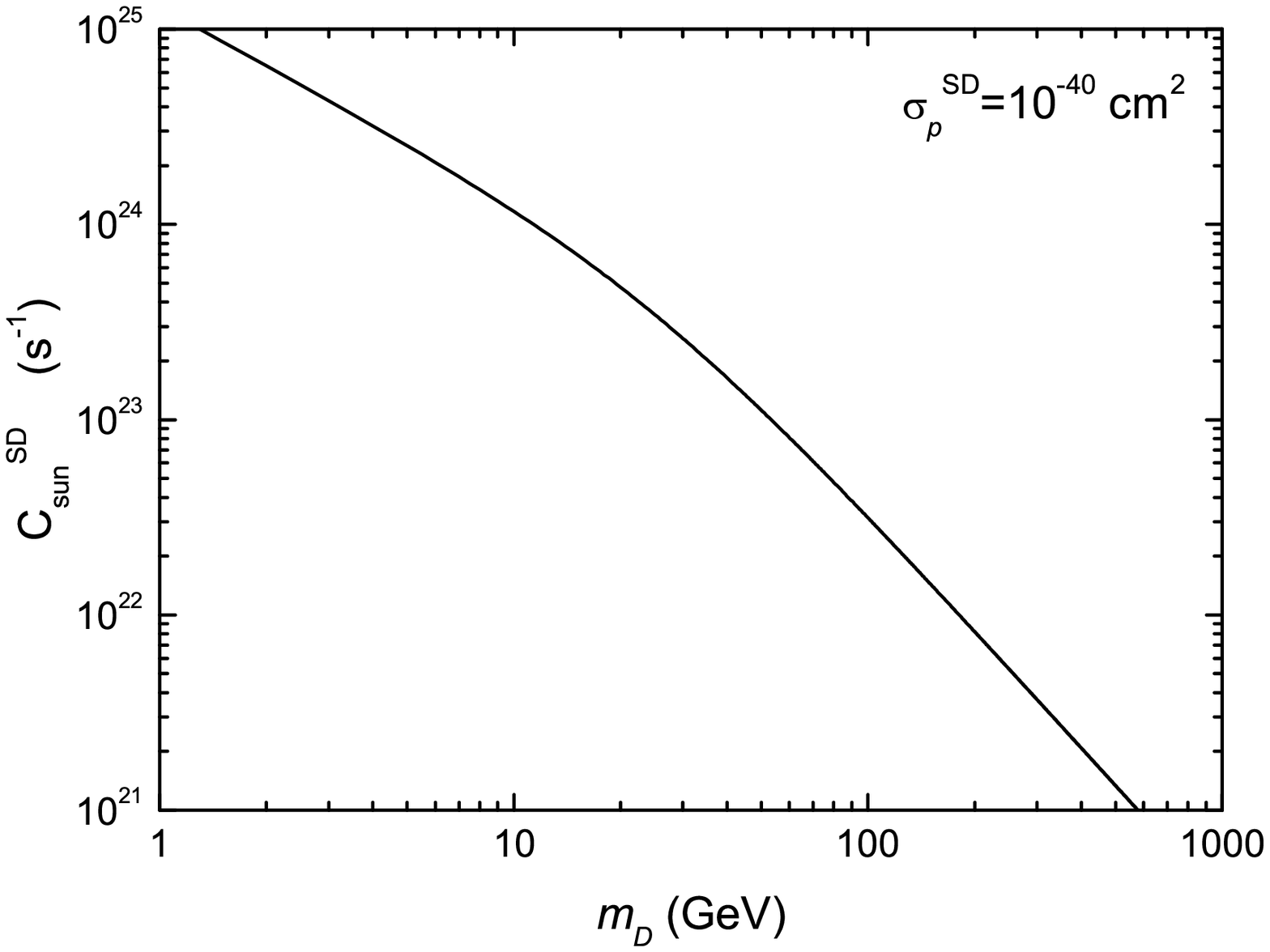}
\end{center}
\vspace{-1.0cm}\caption{ The DM solar capture rates $C^{\rm
SI}_\odot$ (left) and $C^{\rm SD}_\odot$ (right) as a function of
$m_D$ with $\sigma_n^{\rm SI}=\sigma_p^{\rm SD} = 10^{-40}$ cm$^2$.
The related contributions to $C^{\rm SI}_\odot$ from $^1$H, $^4$He,
$^{16}$O, $^{56}$Fe and other 6 elements have also been shown. }
\label{Csun}
\end{figure}

In the following parts, we shall use $C^{\rm SI}_\odot$ and $C^{\rm
SD}_\odot$ to represent the solar capture rate $C_\odot$ in the SI
and SD interaction cases, respectively. $C^{\rm SI}_\odot$ and
$C^{\rm SD}_\odot$ may be approximately written as \cite{DM1}
\begin{eqnarray}
C^{\rm SI}_\odot & \approx & 4.8 \times 10^{24} {\rm s}^{-1}
\frac{\rho_0}{0.3 \, {\rm GeV / cm^3}} \frac{270\, {\rm
km/s}}{\bar{v}} \frac{1 {\rm GeV}}{m_D} \sum_i F_i(m_D)
\frac{\sigma_{{{N}_i}}^{\rm SI}}{10^{-40} {\rm cm^2}} f_i \phi_i
S\left( \frac{m_D}{m_{{N}_i}} \right) \frac{1 {\rm GeV}}{m_{{ N}_i}}
\;,  \label{Csun_SI} \\
 C^{\rm SD}_\odot & \approx & 1.3 \times 10^{25} {\rm s}^{-1}
\frac{\rho_0}{0.3 \, {\rm GeV / cm^3}} \frac{270\, {\rm
km/s}}{\bar{v}} \frac{1 {\rm GeV}}{m_D} \frac{\sigma_{p}^{\rm
SD}}{10^{-40} {\rm cm^2}}  S\left( \frac{m_D}{m_p} \right) \;,
\label{Csun_SD}
\end{eqnarray}
where the local DM density $\rho_0 = 0.3 \, {\rm GeV / cm^3}$ and
the local DM root-mean-square velocity  $\bar{v} = 270 \, {\rm
km/s}$. The function $S(x)$ denotes the kinematic suppression and is
given by
\begin{eqnarray}
S(x) = \left[ \frac{A(x)^{1.5}}{1+A(x)^{1.5}} \right]^{2/3} {\rm
with} \;\; A(x) = \frac{3x}{2(x-1)^2} \left( \frac{\langle v_{\rm
esc} \rangle}{\bar{v}} \right)^2\;,
\end{eqnarray}
where $\langle v_{\rm esc} \rangle = 1156$ km s$^{-1}$ is a mean
escape velocity. $f_i$ and $\phi_i$ describe the mass fraction and
the distribution of the element $N_i$ in the Sun, respectively.
$f_i$, $\phi_i$ and the form-factor suppression $F_i(m_D)$ can be
found in Ref. \cite{DM1}. We shall sum over the following elements
in the Sun: $^1$H, $^4$He, $^{12}$C, $^{14}$N, $^{16}$O, $^{20}$Ne,
$^{24}$Mg, $^{28}$Si, $^{32}$S and $^{56}$Fe. The SI DM-nucleus
elastic scattering cross section $\sigma_{{{ N}_i}}^{\rm SI}$ is
related to the SI DM-nucleon elastic scattering cross section
$\sigma_{n}^{\rm SI}$ by the formula \cite{Guo:2010hq}
\begin{eqnarray}
\sigma_{{N}_i}^{\rm SI} = A_{{ N}_i}^2 \frac{M^2({ N}_i)}{M^2({ n})}
\sigma_{n}^{\rm SI}\;,
\end{eqnarray}
where $A_{{N}_i}$ is the mass number of the nucleus ${N}_i$ and
$M(x) = m_D m_x / (m_D + m_x )$. Here we have assumed that the DM
couplings to protons and neutrons are isospin-invariant.  Assuming
$\sigma_n^{\rm SI} = \sigma_p^{\rm SD} = 10^{-40}$ cm$^2$, we
calculate $C^{\rm SI}_\odot$ and $C^{\rm SD}_\odot$ as shown in Fig.
\ref{Csun}. It is found that  $C^{\rm SI}_\odot$ from other elements
in the Sun is much larger than that from the hydrogen element
although it has the maximal mass fraction. In the SD case, the
hydrogen element plays the dominant role.

Considering the usual $\langle \sigma v \rangle \approx 3 \times
10^{-26}$ cm$^3$ s$^{-1}$ induced from the observed DM relic
density, we can obtain $C_\odot \geq 8.6 \times 10^{22}/(m_D/1{\rm
GeV})^{3/2} {\rm s}^{-1}$ \cite{Guo:2013ypa} from $t_\odot \sqrt{
C_\odot C_A} \geq 3.0$ which means $\tanh^2[t_\odot \sqrt{ C_\odot
C_A}] \geq 0.99$ in Eq. (\ref{GammaA}). In terms of the results in
Fig. \ref{Csun}, we may safely assume that the DM annihilation rate
reaches equilibrium with the DM capture rate. Then the electron
neutrino flux at the surface of the Earth from the solar DM
annihilation can be written as:
\begin{eqnarray}
\frac{d \Phi_{\nu_{e}} }{d E_{\nu} } = \frac{\Gamma_{A}}{4 \pi
R_{\rm ES}^2} \frac{d N_{\nu_{e}} }{d E_{\nu} } \simeq
\frac{C_\odot}{8 \pi R_{\rm ES}^2} \frac{d N_{\nu_{e}} }{d E_{\nu} }
\;, \label{dphide}
\end{eqnarray}
where $R_{\rm ES} = 1.496 \times 10^{13}$ cm is the Earth-Sun
distance. $d N_{\nu_{e}}/d E_{\nu} $ is the differential electron
neutrino energy spectrum  from per DM pair annihilation in the Sun.
In order to calculate  $d N_{\nu_e}/d E_{\nu}$,  one must consider
the hadronization, interactions and decay processes of the DM
annihilation final states in the core of the Sun.  In addition, we
should consider the neutrino interactions in the Sun and neutrino
oscillations. Here the following neutrino oscillation parameters
will be chosen as input values: \cite{Capozzi:2013csa,An:2012eh}
\begin{eqnarray}
\sin^2 \theta_{12} & = & 0.308 ,\;\; \sin^2 \theta_{23} = 0.437,
\;\; \sin^2 \theta_{13} = 0.0234,  \nonumber \\
\Delta m_{21}^2  &  = &  7.54 \times 10^{-5} {\rm eV}^2, \;\; \Delta
m_{31}^2 = 2.47 \times 10^{-3} {\rm eV}^2, \;\; \delta=0^\circ.
\label{osci}
\end{eqnarray}
Then we use the program package WimpSim \cite{WimpSim} to calculate
$d N_{\nu_e}/d E_{\nu}$ and  $d N_{\bar{\nu}_e}/d E_{\nu}$ for three
typical DM annihilation channels $\chi \chi \rightarrow \nu
\bar{\nu}, \tau^+ \tau^-, b \bar{b}$ and $4 \; {\rm GeV} \leq m_D
\leq 20$ GeV. Note that $\nu_e \bar{\nu}_e$, $\nu_\mu \bar{\nu}_\mu$
and $\nu_\tau \bar{\nu}_\tau$ have the same contributions in the
$\nu \bar{\nu}$ channel.

\section{Electron neutrinos and antineutrinos in JUNO}

Here we shall discuss the $\nu_e/\bar{\nu}_e$ signals in JUNO from
the DM $\nu \bar{\nu}$, $\tau^+ \tau^-$ and $b \bar{b}$ annihilation
channels. The JUNO central detector holds 20 kton LS which will be
in a spherical container of radius of 17.7 m \cite{JUNO}. There is
1.5 m water buff region between about 17000 20-inch PMTs and the LS
surface. According to the detector properties, we have made a MC
simulation based on the GENIE generator \cite{GENIE} and the Geant4
detector simulation \cite{GEANT4}. For the $\nu_e/\bar{\nu}_e$
charged current (CC) interactions in the JUNO detector, the MC
simulation can provide many useful information which includes the
event visible energy $E_{vis}$, the $e^\pm$ visible energy
$E^e_{vis}$, the initial neutrino direction, the final state $e^\pm$
direction, etc.

The JUNO can reconstruct $E_{vis}$ with a very excellent energy
resolution $\sigma_{E_{vis}}$. For the $\nu_e/\bar{\nu}_e$ from
$4-20$ GeV DM annihilation, we may conservatively take
$\sigma_{E_{vis}} = 0.01 \sqrt{E_{vis}/{\rm GeV}}$  which origins
from the statistical fluctuation in the scintillation photon
emission and the quenching fluctuation \cite{JUNO}. The
$\sigma_{E_{vis}}$ will be neglected in the following analysis since
it has the very slight effects. On the other hand, the JUNO can also
reconstruct the single muon direction with the angular resolution
better than $1^\circ$ if its track length $L_\mu> 5$ m and intrinsic
PMT timing resolution better than 4 ns \cite{JUNO}. For the single
electron/positron track, the 50 kton LS detector LENA find that the
angular resolution is a few degrees \cite{LENA}. Note that the
hadronic final states in the $\nu_e$ and $\bar{\nu}_e$ CC
interactions will affect the $e^\pm$ angular resolution and the
identification of electron shower. The related studies are under
way. Here we assume the $\nu_e/\bar{\nu}_e$ CC events with
$E^e_{vis} > 1$ GeV and $Y_{vis} \equiv E^e_{vis}/E_{vis} > 0.5$ can
be identified and reconstructed very well. For these selected
events, the $e^\pm$ angular resolution will be assumed to be
$10^\circ$ in the following analysis.

\begin{figure}[htb]
\begin{center}
\includegraphics[scale=0.46]{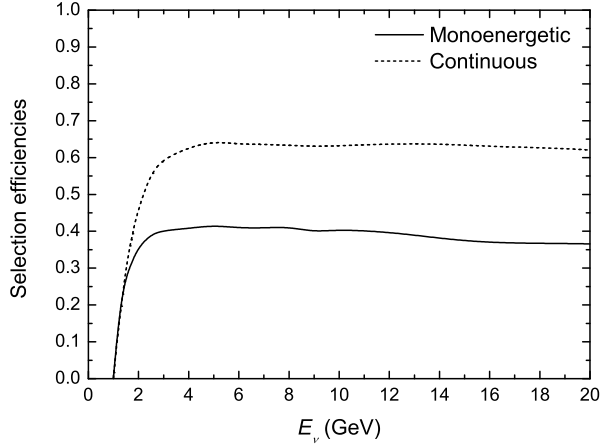}
\end{center}
\vspace{-0.0cm}\caption{ The $\nu_e/\bar{\nu}_e$ CC event selection
efficiencies $\epsilon(E_{\nu})$ as a function of initial neutrino
energy $E_\nu$ based on the corresponding selection conditions in
the monoenergetic and continuous spectrum cases. } \label{eff}
\end{figure}

For the DM $\nu \bar{\nu}$ annihilation channel, the WimpSim gives
an approximate monoenergetic $\nu_e/\bar{\nu}_e$ spectrum. This is
because that the initial monoenergetic $\nu_e/\bar{\nu}_e$ spectrum
($E_\nu = m_D$) will be slightly modified by the neutrino
interactions in the Sun. The DM $\tau^+ \tau^-$ and $b \bar{b}$
annihilation channels have the continuous spectra. In order to
suppress the atmospheric neutrino backgrounds, we shall take
different selection conditions for a given $\nu_e/\bar{\nu}_e$
energy $E_\nu$ in the monoenergetic and continuous
$\nu_e/\bar{\nu}_e$ spectrum cases. Then two selection efficiencies
$\epsilon(E_{\nu})$ will be obtained from the MC simulation data.
Except for $E^e_{vis} > 1$ GeV and $Y_{vis} \equiv E^e_{vis}/E_{vis}
> 0.5$, we only select the events with $\theta_{\rm  sun} < 20^\circ
\sqrt{10 {\rm GeV}/E_\nu}$ \cite{DM1} and $1 \geq E_{vis}/E_\nu
> 0.9$ for the monoenergetic spectrum case. Here $\theta_{\rm sun}$
denotes the angle between the initial neutrino direction (the Sun
direction) and the reconstructed $e^\pm$ direction. Based on the
$e^\pm$ initial direction and $10^\circ$ angular resolution, one can
easily calculate the angle $\theta_{\rm sun}$  for every MC
simulation event. For the continuous spectrum case,  $E^e_{vis} > 1$
GeV, $Y_{vis} > 0.5$, $\theta_{\rm sun} < 30^\circ $ and $ 1 \, {\rm
GeV} < E_{vis} < E_\nu$ will be chosen. Then we calculate the
corresponding selection efficiencies $\epsilon(E_{\nu})$  for a
given energy $\nu_e/\bar{\nu}_e$ in the monoenergetic and continuous
spectrum cases as shown in Fig. \ref{eff}. It is found that
$\epsilon(E_{\nu})$ will not obviously change for $E_\nu > 3$ GeV.

\begin{figure}[htb]
\begin{center}
\includegraphics[scale=0.46]{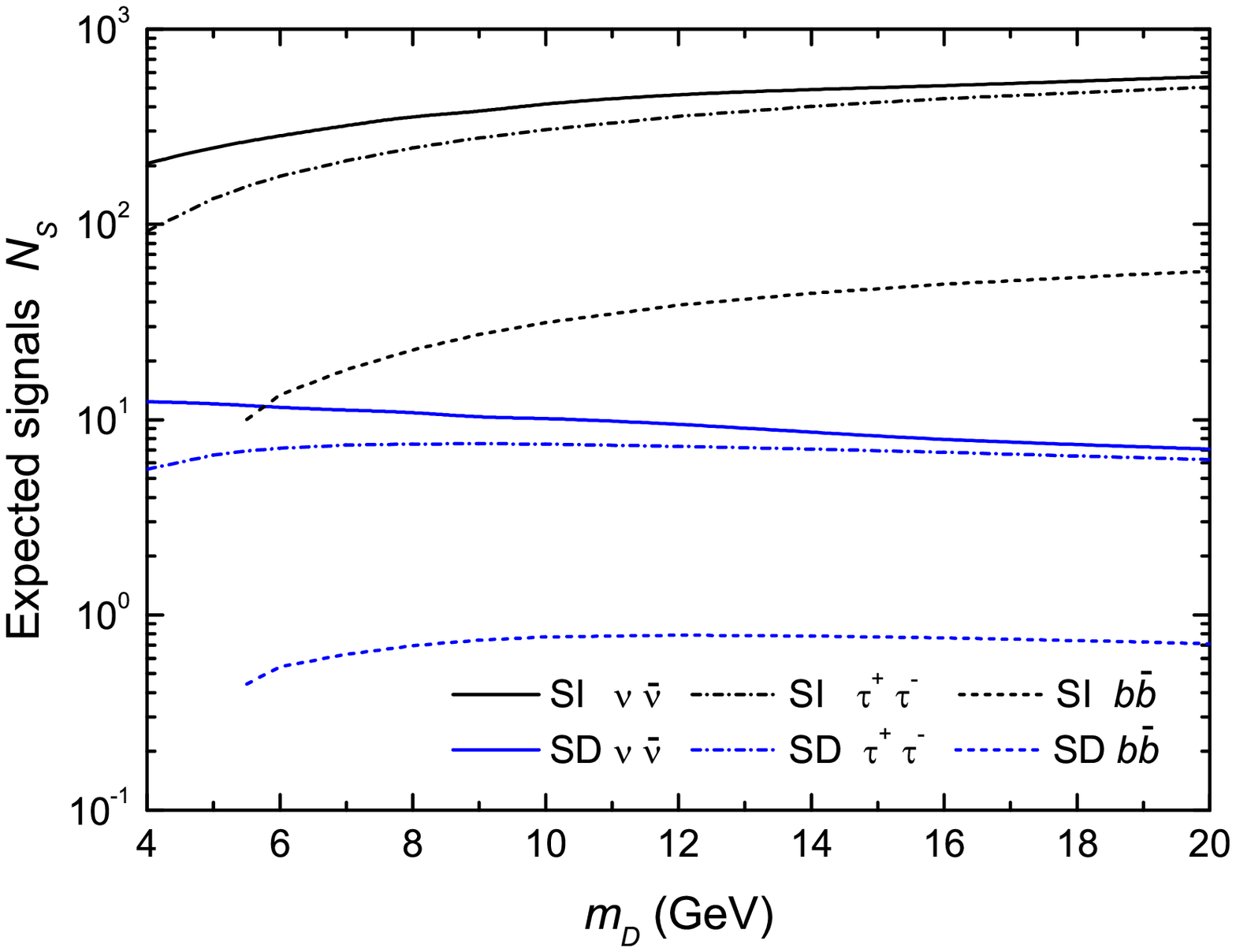}
\includegraphics[scale=0.46]{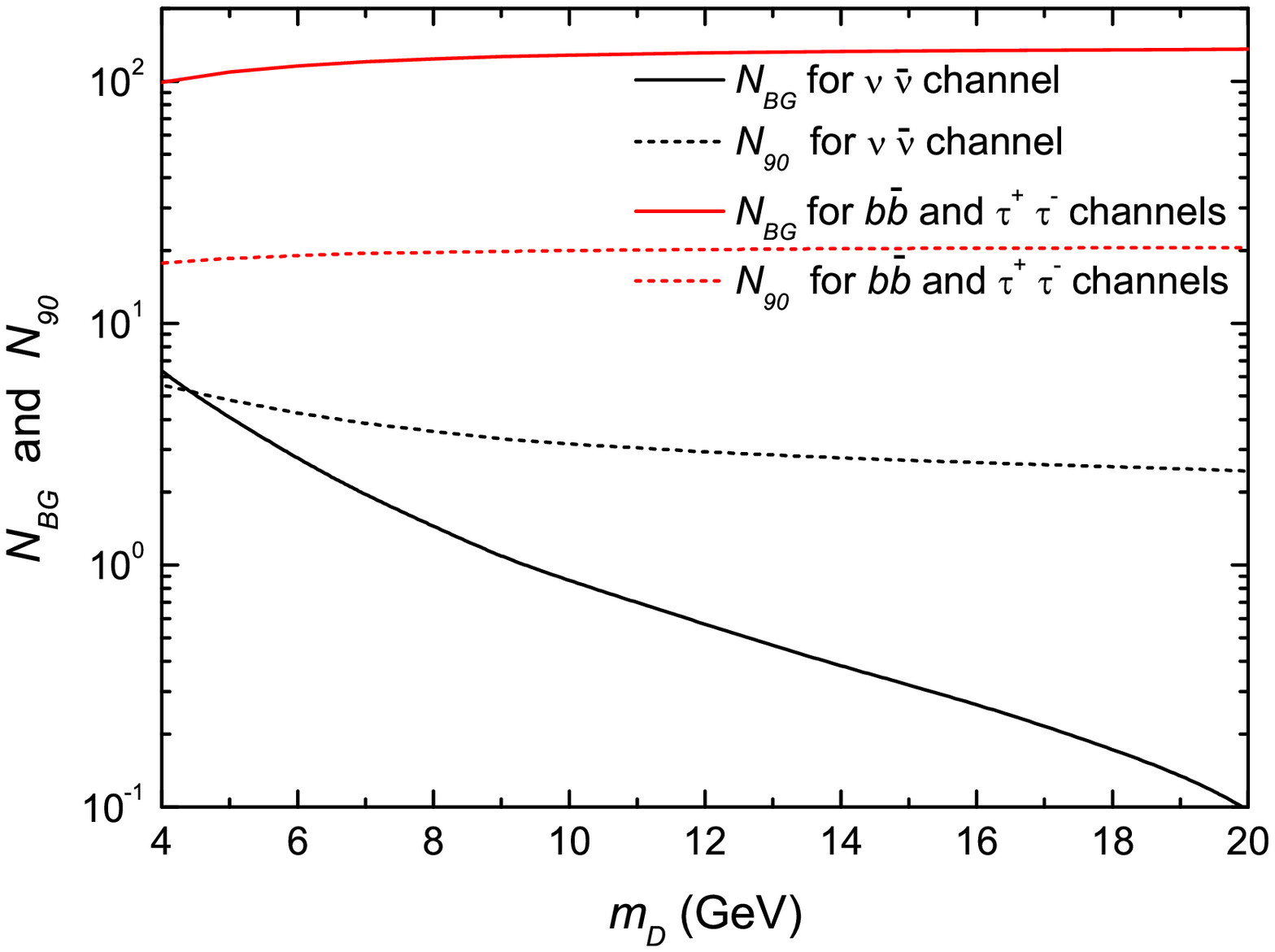}
\end{center}
\vspace{-1.0cm}\caption{Left panel:  the expected signals $N_S$ from
the DM $\nu \bar{\nu}$, $\tau^+ \tau^-$ and $b \bar{b}$ channels for
the $\sigma_n^{\rm SI} = 10^{-40}$ cm$^2$ and $\sigma_p^{\rm SD} =
10^{-40}$ cm$^2$ cases. Right panel: the atmospheric
$\nu_e/\bar{\nu}_e$ CC backgrounds $N_{BG}$ and the deduced 90\% CL
upper limit $N_{90}$ to $N_S$ for the monoenergetic and continuous
spectrum cases.} \label{NsNbgN90}
\end{figure}

For a given DM mass $m_D$, the expected $\nu_e/\bar{\nu}_e$ CC event
numbers from  the DM annihilation in the Sun can be expressed as
\begin{eqnarray}
N_S =  N_n \, t  \int^{m_D}_{E_{th}} \left[\frac{d \Phi_{\nu_e} }{d
E_{\nu} } \sigma_{\nu_e} + \frac{d \Phi_{\bar{\nu}_e} }{d E_{\nu} }
\sigma_{\bar{\nu}_e} \right] \epsilon(E_{\nu}) \, d E_{\nu} \;,
\label{NS}
\end{eqnarray}
where the total nucleon number $N_n \simeq 20\, {\rm kton} / m_n$
and $m_n$ is the nucleon mass. In the following parts, we shall take
the JUNO exposure time $t = 10$ years. Here we extract ${\nu_e}$
($\bar{\nu}_e$) per nucleon CC cross section $\sigma_{\nu_e}$
($\sigma_{\bar{\nu}_e}$) from the GENIE \cite{GENIE} and consider
that the LS target includes 12\% $^1$H and 88\% $^{12}$C.
$\sigma_{\nu_e}$ and $\sigma_{\bar{\nu}_e}$ can also be found in the
left panel of Fig. 7-5 in Ref. \cite{JUNO}. In Eq. (\ref{NS}), the
integral lower limit $E_{th} = 0.9 m_D$ ($E_{th} = 1$ GeV) for the
DM $\nu \bar{\nu}$ channel ($\tau^+ \tau^-$ and $b \bar{b}$
channels) will be chosen. With the help of Eq. (\ref{dphide}), Eq.
(\ref{NS}) and the efficiencies $\epsilon(E_{\nu})$ in Fig.
\ref{eff}, one can calculate the expected event numbers $N_S$ for
different DM masses and annihilation channels. For illustration, we
plot $N_S$ as a function of $m_D$ in the left panel of Fig.
\ref{NsNbgN90} with $\sigma_n^{\rm SI} = 10^{-40}$ cm$^2$ and
$\sigma_p^{\rm SD} = 10^{-40}$ cm$^2$.

\section{Sensitivities to the DM direct detection cross sections }

For the $\nu_e/\bar{\nu}_e$ signals from the DM annihilation in the
Sun, the related backgrounds are due to the atmospheric neutrino CC
and neutral current (NC) interactions in the JUNO. The atmospheric
$\nu_e/\bar{\nu}_e$ CC events are the irreducible background. The
atmospheric neutrino NC background mainly origins from the $\pi^0$
misidentification as $e^\pm$. As shown in Fig. 7-5 of Ref.
\cite{JUNO}, the NC event rate is about half of the
$\nu_e/\bar{\nu}_e$ CC event rate for $E_{vis}>1$ GeV. Note that the
$e^\pm$ takes on average about $60\%$ of the initial
$\nu_e/\bar{\nu}_e$ energy in the CC interaction. For the NC
interaction, several hadronic particles will usually share
$E_{vis}$. Considering the requirement  $E^e_{vis} > 1$ GeV and the
$\pi^0$ misidentification rate, we shall neglect the atmospheric
neutrino NC background and only calculate the atmospheric
$\nu_e/\bar{\nu}_e$ CC background.

With the help of the GENIE generator and Geant4 detector simulation,
1.5 million $\nu_e/\bar{\nu}_e$ CC events in JUNO detector have been
simulated \cite{JUNO}. On the other hand, we calculate the expected
atmospheric $\nu_e/\bar{\nu}_e$ CC event numbers in every energy and
zenith angle bin by use of the atmospheric neutrino fluxes
\cite{Honda:2015fha} and the oscillation parameters in Eq.
(\ref{osci}). Comparing the event numbers per bin of the MC
simulation and the corresponding theoretical values, we determine
the weight value for every MC event. Then the expected atmospheric
$\nu_e/\bar{\nu}_e$ CC sample can be obtained. For the DM $\nu
\bar{\nu}$ channel, we apply the selection conditions $E^e_{vis} >
1$ GeV, $Y_{vis}> 0.5$ and $1 \geq E_{vis}/m_D > 0.9$ to calculate
the corresponding atmospheric $\nu_e/\bar{\nu}_e$ CC event numbers
from all directions. Then we average the all direction result and
obtain the atmospheric $\nu_e/\bar{\nu}_e$ CC background $N_{BG}$
within the cone half-angle $\theta_{\rm sun} < 20^\circ \sqrt{10
{\rm GeV}/m_D}$. For the $\tau^+ \tau^-$ and $b \bar{b}$ channels,
$E^e_{vis} > 1$ GeV, $Y_{vis}> 0.5$, $1 \, {\rm GeV} < E_{vis} <
m_D$ and $\theta_{\rm sun} < 30^\circ$ will be used. In the right
panel of Fig. \ref{NsNbgN90}, we plot the atmospheric
$\nu_e/\bar{\nu}_e$ CC background $N_{BG}$ as a function of $m_{D}$.
It is clear that the $\nu \bar{\nu}$ channel has the smaller
$N_{BG}$ than the $\tau^+ \tau^-$ and $b \bar{b}$ channels. This is
because that  we take  $1 \, {\rm GeV} < E_{vis} < m_D$ and a larger
cone half-angle $\theta_{\rm sun}$ for the DM $\tau^+ \tau^-$ and $b
\bar{b}$ channels.

To estimate the JUNO sensitivities to the direct detection cross
sections $\sigma_n^{\rm SI}$ and $\sigma_p^{\rm SD}$, we assume the
observed event number $N_{obs} = N_{BG}$. Then the 90\% confidence
level (CL) upper limit $N_{90}$ to the expected $\nu_e/\bar{\nu}_e$
signals $N_{S}$ can be derived through the following formulas
\cite{SK}:
\begin{eqnarray}
90 \%  = \frac{\int_{N_{S} = 0}^{N_{90}} L(N_{ obs}|N_{S}) d N_{
S}}{\int_{N_{ S} = 0}^{\infty} L(N_{ obs}|N_{ S}) d N_{S}}
\label{N90}
\end{eqnarray}
with the Poisson-based likelihood function
\begin{eqnarray}
L(N_{obs}|N_S) =  \frac{(N_S + N_{BG})^{N_{obs}}}{N_{obs}!} e^{-(N_S
+ N_{BG})} \;.
\label{L}
\end{eqnarray}
In terms of $N_{BG}$, we calculate the corresponding upper limit
$N_{90}$ for the $\nu \bar{\nu}$, $\tau^+ \tau^-$ and $b \bar{b}$
channels. In the right panel of Fig. \ref{NsNbgN90}, we plot
$N_{90}$ as a function of $m_D$. It is found that $N_{90}$ decreases
from 5.6 to 2.5 as the DM mass $m_D$ increases in the $\nu
\bar{\nu}$ channel case. For the $\tau^+ \tau^-$ and $b \bar{b}$
channels, $N_{90}$ only slightly changes.

\begin{figure}[htb]
\begin{center}
\includegraphics[scale=0.46]{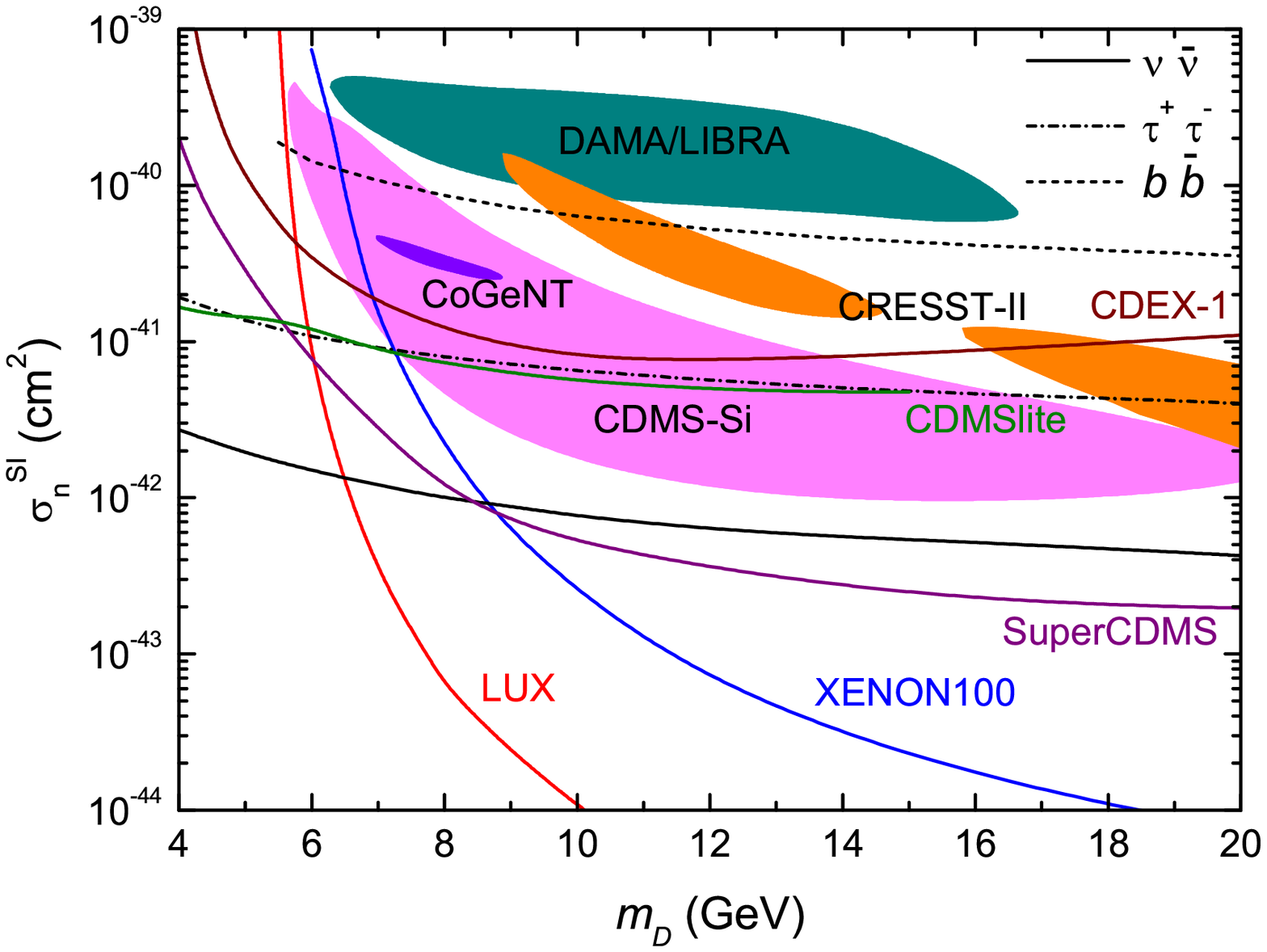}
\includegraphics[scale=0.46]{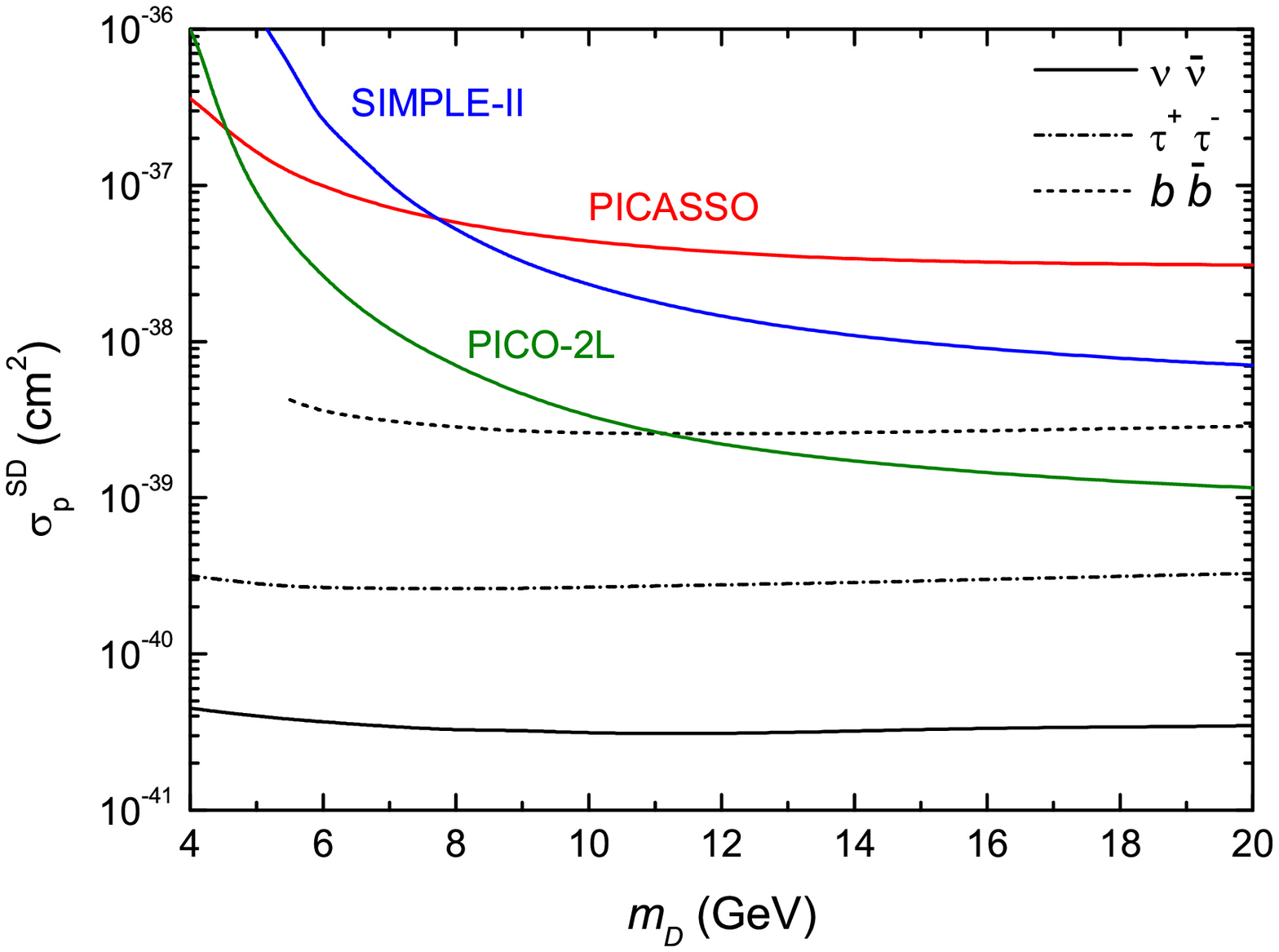}
\end{center}
\vspace{-1.0cm}\caption{ The JUNO sensitivities to $\sigma_n^{\rm
SI}$ (left) and $\sigma_p^{\rm SD}$ (right) as a function of $m_D$
for the DM $\nu \bar{\nu}$, $\tau^+ \tau^-$ and $b \bar{b}$
annihilation channels with 10 year data.} \label{limits}
\end{figure}

The JUNO sensitivities to $\sigma_n^{\rm SI}$ and $\sigma_p^{\rm
SD}$ can be derived from $N_S = N_{90}$  and Eq. (\ref{NS}). In Fig.
\ref{limits}, we plot the JUNO sensitivities as a function of $m_D$
for the DM $\nu \bar{\nu}$, $\tau^+ \tau^-$ and $b \bar{b}$ channels
with 10 year running. Here the current experimental results from the
direct detection experiments have also been shown. Note that
DAMA/LIBRA \cite{DAMA,Savage:2008er}, CDMS-Si \cite{CDMS-Si}, CoGeNT
\cite{CoGeNT} and CRESST-II \cite{CRESST,Brown:2011dp} have the
positive results which can be interpreted as arising from the SI
DM-nucleon interaction with the shadowed parameter space
\cite{CDMS-Si} as shown in the left panel of Fig. \ref{limits}.
However the LUX \cite{LUX}, XENON100 \cite{XENON100}, CDMSlite
\cite{CDMSlite}, SuperCDMS \cite{SuperCDMS} and CDEX-1
\cite{Yue:2014qdu} experiments do not find the DM evidence and only
give the upper limits to $\sigma^{\rm SI}_n$. It is clear that the
JUNO projected sensitivity to $\sigma^{\rm SI}_n$ is better than the
direct detection experiments for the $\nu \bar{\nu}$ channel with
$m_D< 6.5$ GeV. If the $\nu \bar{\nu}$ or $\tau^+ \tau^-$ channel is
dominant, the future JUNO results are very helpful for us to
understand the tension between the DAMA annual modulation signal and
other direct detection exclusions. In the SD case, the JUNO
sensitivities to $\sigma_p^{\rm SD}$ from the $\nu \bar{\nu}$ and
$\tau^+ \tau^-$ channels are much better than the current PICO-2L
\cite{PICO}, PICASSO \cite{PICASSO} and SIMPLE-II \cite{Simple}
experimental limits.

\section{Conclusions}

In conclusion, we have investigated the $\nu_e$ and $\bar{\nu}_e$ CC
signals in JUNO from the DM annihilation in the Sun. Three typical
DM annihilation channels $\chi \chi \rightarrow \nu \bar{\nu},
\tau^+ \tau^-, b \bar{b}$ have been analyzed for the light DM mass
$4 \; {\rm GeV} \leq m_D \leq 20$ GeV. In terms of the spectrum
features, we take two sets of selection conditions for the
monoenergetic ($\nu \bar{\nu}$ channel) and continuous ( $\tau^+
\tau^-$ and $b \bar{b}$ channels) spectra. The corresponding
selection efficiencies $ \epsilon (E_\nu)$ can be derived for a
given $\nu_e/\bar{\nu}_e$ energy from the MC simulation data and
$10^\circ$ angular resolution. Then we numerically calculate the
expected $\nu_e$ and $\bar{\nu}_e$ CC event numbers for the SI and
SD interactions. On the other hand, we calculate the irreducible
atmospheric $\nu_e/\bar{\nu}_e$ CC background which is far larger
than the atmospheric neutrino NC background. Finally, we present the
JUNO sensitivities to $\sigma_n^{\rm SI}$ and $\sigma_p^{\rm SD}$
for the DM $\nu \bar{\nu}$, $\tau^+ \tau^-$ and $b \bar{b}$
annihilation channels with 10 year running. It is found that the
JUNO projected sensitivities to $\sigma_p^{\rm SD}$ are much better
than the current experimental upper limits for the $\nu \bar{\nu}$
and $\tau^+ \tau^-$ channels. For the SI case, the JUNO will give
the better sensitivity to $\sigma_n^{\rm SI}$ than the current
direct detection experiments for $m_D< 6.5$ GeV in the $\nu
\bar{\nu}$ channel case.

\acknowledgments

I am grateful to Tao Lin and Jia-Shu Lu for their useful help in the
Geant4 detector simulation and the GENIE generator. This work is
supported in part by the National Nature Science Foundation of China
(NSFC) under Grants No. 11575201  and the Strategic Priority
Research Program of the Chinese Academy of Sciences under Grant No.
XDA10010100.

\end{document}